\documentclass[showpacs,aps,prl,twocolumn,superscriptaddress]{revtex4-1} 
\usepackage[pdftex]{graphicx} 
\usepackage{amsmath} 
\usepackage{amsfonts} 
\usepackage{amssymb} 
\usepackage{amsbsy} 
\usepackage{ulem} 
\usepackage{color} 
\usepackage{times} 
\usepackage{units}

\begin{document}

\title{Quantum Hall Effect and Semimetallic Behavior of Dual-Gated ABA-Stacked Trilayer Graphene} 

\author{E. A. Henriksen} 
\altaffiliation{Electronic address: erikku@caltech.edu} 
\affiliation{Condensed Matter Physics, California Institute of Technology, Pasadena, CA 91125}  

\author{D. Nandi}
\affiliation{Condensed Matter Physics, California Institute of Technology, Pasadena, CA 91125}

\author{J. P. Eisenstein} 
\affiliation{Condensed Matter Physics, California Institute of Technology, Pasadena, CA 91125}  

\begin{abstract}
The electronic structure of multilayer graphenes depends strongly on the number of layers as well as the stacking order. Here we explore the electronic transport of purely ABA-stacked trilayer graphenes in a dual-gated field-effect device configuration. We find that both the zero-magnetic-field transport and the quantum Hall effect at high magnetic fields are distinctly different from the monolayer and bilayer graphenes, and that they show electron-hole asymmetries that are strongly suggestive of a semimetallic band overlap. When the ABA trilayers are subjected to an electric field perpendicular to the sheet, Landau level splittings due to a lifting of the valley degeneracy are clearly observed.
\end{abstract}

\date{\today}


\maketitle 

Single and bilayer graphenes are well known to have low-energy electronic dispersions that differ dramatically from the parent compound, graphite \cite{castroneto:109,abergel:261,dassarma:407}. With just one more layer, two stable forms of trilayer graphene are known to exist having either ABA (Bernal)- or ABC (rhombohedral)-stacking, which exhibit differing electronic structures. In particular, ABA-stacked trilayer graphene is predicted to be a semimetal and is thus the first-- and thinnest-- multilayer graphene to have a band structure resembling that of bulk graphite \cite{latil:036803,guinea:245426,partoens:075404,aoki:123,koshino:125443,brandt}. Not only do trilayer graphenes comprise a new class of materials intermediate between graphite and its fundamental building block of monolayer graphene, but studies of the electronic transport of ABA graphenes may also shed light on transport anomalies observed in bulk graphite and other semimetals whose properties cannot be as easily varied by externally applied gate voltages \cite{yaguchi:344207,behnia:1729}.

Trilayer graphene has received increasing attention of late, due to interest in the unusual electronic structures of ABA- and ABC-stacked graphene, and also to advances in identifying the nature of the stacking order of trilayer flakes. The two crystal structures differ in that for ABA graphene the top and bottom sheets are aligned while the middle sheet is shifted by one bond length along the C-C bond direction (see Fig. 1 (a) ); while for ABC graphene the top layer is also shifted, by two bond lengths relative to the bottom layer. Measurements of the electronic transport in trilayers have been previously reported \cite{craciun:383}, and the nature of the stacking has been inferred by comparing the observed signatures of the quantum Hall effect (QHE) to those expected for ABA \cite{taychatanapat:1} and ABC \cite{kumar:126806} graphenes. However, the recent development of optical methods to clearly distinguish between regions of differing stacking order has lead to investigations of the electronic properties of trilayers known to be chiefly of one type \cite{lui:164,lui:1,bao:1,zhang:1,jhang:4995}. In this work, we study large area dual-gated trilayer graphene samples known to be of ABA stacking and present a unified view of the electronic transport of this system that has not been available to date. We find that the zero-, low-, and high-magnetic-field quantum Hall transport of ABA-stacked trilayer graphene samples can be understood in the context of the underlying semimetallic band structure that arises from the mirror-symmetric ABA crystal lattice.

On average, trilayer graphene contains regions of ABA- and ABC-stacking in an 85:15 ratio. The two forms pose a problem for measurements of the QHE where the nature of the edge states may change, depending on the stacking type, to such an extent that even a small region of differing stacking order can affect the interpretation of transport data. Therefore in this work only those graphene flakes identified as being predominantly composed of ABA-stacked trilayers by Fourier-transform infrared (IR) spectroscopy are utilized. Moreover as for bilayer graphene, in multilayer graphenes the full range of electronic transport cannot be accessed by a single-side gated device, since varying the voltage on just one gate simultaneously changes both the Fermi level (or carrier density $n$) and the drop in electric potential (or electric displacement field, $D$) between the layers \cite{zhang:820,henriksen:041412,avetisyan:195401}. This problem is remedied by the use of a second (top) gate which allows for independent control over the values of $n$ and $D$. In ABA-stacked trilayers we will see it is crucial to distinguish between effects that depend on only one or the other parameter in order to understand the observed electronic transport.

\begin{figure*}[t]
\includegraphics[width=0.9\textwidth]{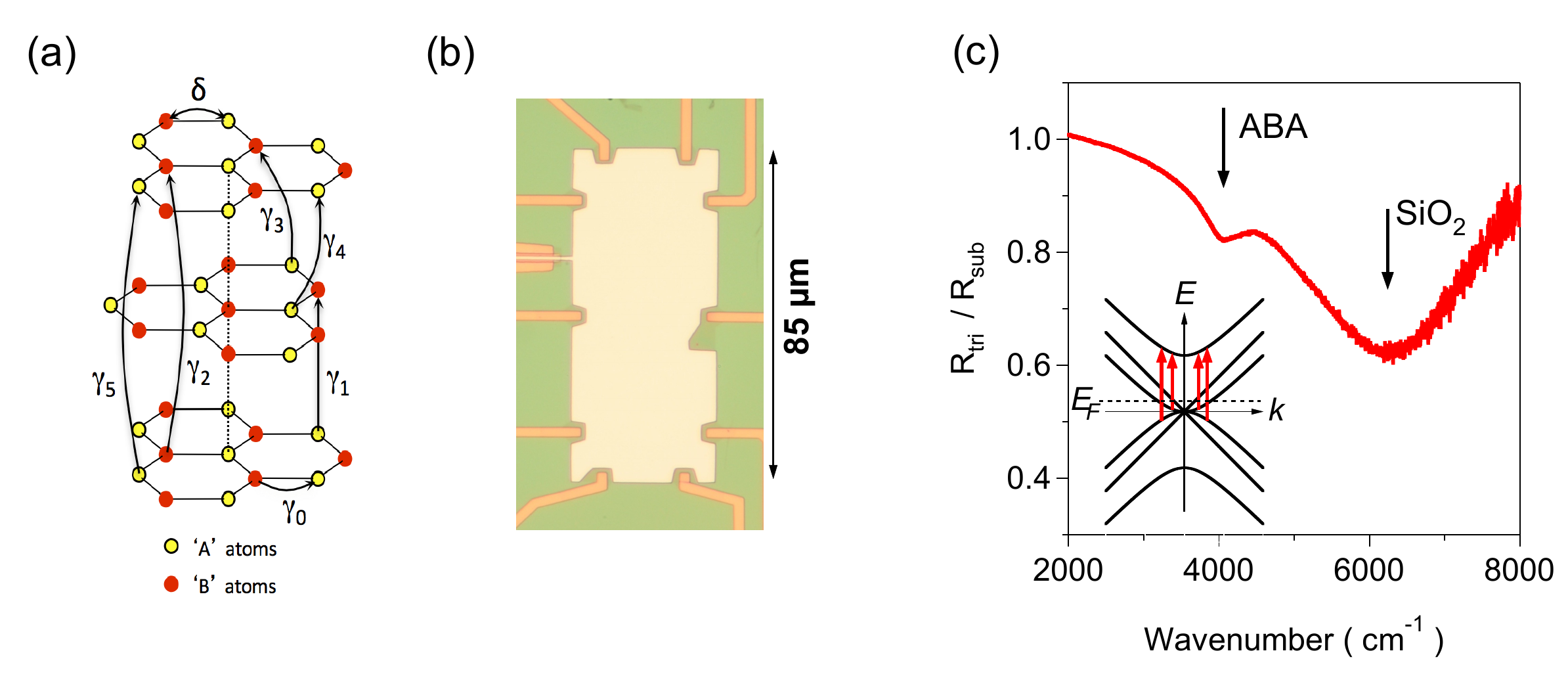}
\caption{(Color) (a) Crystal structure of ABA (Bernal)- stacked trilayer graphene, with arrows indicating the tight-binding hopping parameters of the Slonczewski-Weiss-McClure model \cite{koshino:125443,brandt}. (b) Optical image of a 3000 $\mu$m$^2$ dual-gated ABA trilayer graphene device. (c) Infrared reflectance, $R_\text{tri}$, of the trilayer flake shown in (b) normalized to the substrate reflectance, $R_\text{sub}$. The dip near 4000 cm$^{-1}$ is due to transitions to or from the split off bands, as shown schematically in the lower left inset for the case of light electron doping. The larger dip at 6000 cm$^{-1}$ is due to interference in the thin SiO$_2$ substrate \cite{li:532}.}
\end{figure*}

In Fig. 1 (b) an optical microscope image of a typical device with an area of 3000 $\mu$m$^2$ is shown. To determine the layer stacking, each flake is studied via Fourier-transform IR spectroscopy prior to any device fabrication. The IR reflectance at room temperature, $R_{\text{tri}}$, averaged over the entirety of the graphene sheet embedded in this device and normalized by the reflectance of the nearby substrate, $R_{\text{sub}}$, is shown in Fig. 1 (c). Two features stand out: A broad minimum centered at 6000 cm$^{-1}$ and a smaller dip near 4000 cm$^{-1}$. While the former is due to interference effects in the 285 nm-thick SiO$_2$ layer \cite{li:532}, the latter feature arises from transitions to (from) the split-off conduction (valence) band. The relevant transitions when the system is lightly doped are shown schematically in the inset to Fig. 1 (c). These are similar to the IR absorption in bilayer graphene although higher in energy by a factor of $\sqrt{2}$ \cite{mak:14999,li:037403,kuzmenko:115441}. Along with the absence of any IR signal due to ABC graphene which would appear at 2500-3000 cm$^{-1}$, and with independent confirmation via Raman spectroscopy measurements, the 4000 cm$^{-1}$ dip clearly identifies the flake as ABA-stacked trilayer graphene \cite{lui:164,lui:1}. In particular, analysis of the reflectance data allows an estimate for the content of the ABC-stacked trilayer graphene in this sample at less than 3$\%$.

Graphene-on-SiO$_2$ samples found to have ABA stacking are fabricated into dual-gated field-effect devices via standard electron beam lithography and thin film deposition techniques. The top gate dielectric is composed of polymethyl methacrylate (PMMA) 950 K A4 diluted with methoxybenzene (anisole), and exposed to a 25 kV electron beam at a dosage of $24,000 ~\mu$C/cm$^2$ to create a hardened dielectric layer. Applying voltages $V_\text{b}$ and $V_\text{t}$ to the back and top gates leads to changes in both $n$ and $D$, as given by $n=\alpha \left(\Delta V_\text{b} + \beta \Delta V_\text{t}\right)$ and $D=(D_\text{b} - D_\text{t})/2=(\epsilon_\text{b} \Delta V_\text{b} / d_\text{b} - \epsilon_\text{t} \Delta V_\text{t} / d_\text{t})/2$, with $\Delta V_\text{i} = (V_\text{i} - V_\text{i0})$. Here the $d_\text{i}$ and $\epsilon_\text{i}$ are the thicknesses and dielectric constants of the insulating layers, respectively, and the $V_\text{i0}$ are the offset voltages required to reach $n=D=0$ due to extrinsic doping. Both the density and the gating efficiency, $\alpha=7.5\times10^{10}$ cm$^{-2}$V$^{-1}$, are calibrated by oscillations in the magnetoresistance, $R_{xx}$, at high magnetic fields. The ratio of the two gate capacitances, $\beta$, is determined from the slope, $m$, of the high-resistance ridge in the zero-field resistivity measurements in Fig. 2 (a) by $- m=1/\beta=d_\text{t} \epsilon_\text{b} / (d_\text{b} \epsilon_\text{t})$. The precise location of $D=0$ is determined via features in the QHE, as described below. All electronic transport measurements are made at $T=0.3$ K using standard low-frequency lock-in techniques.

\begin{figure*}[t]
\includegraphics[width=0.9\textwidth]{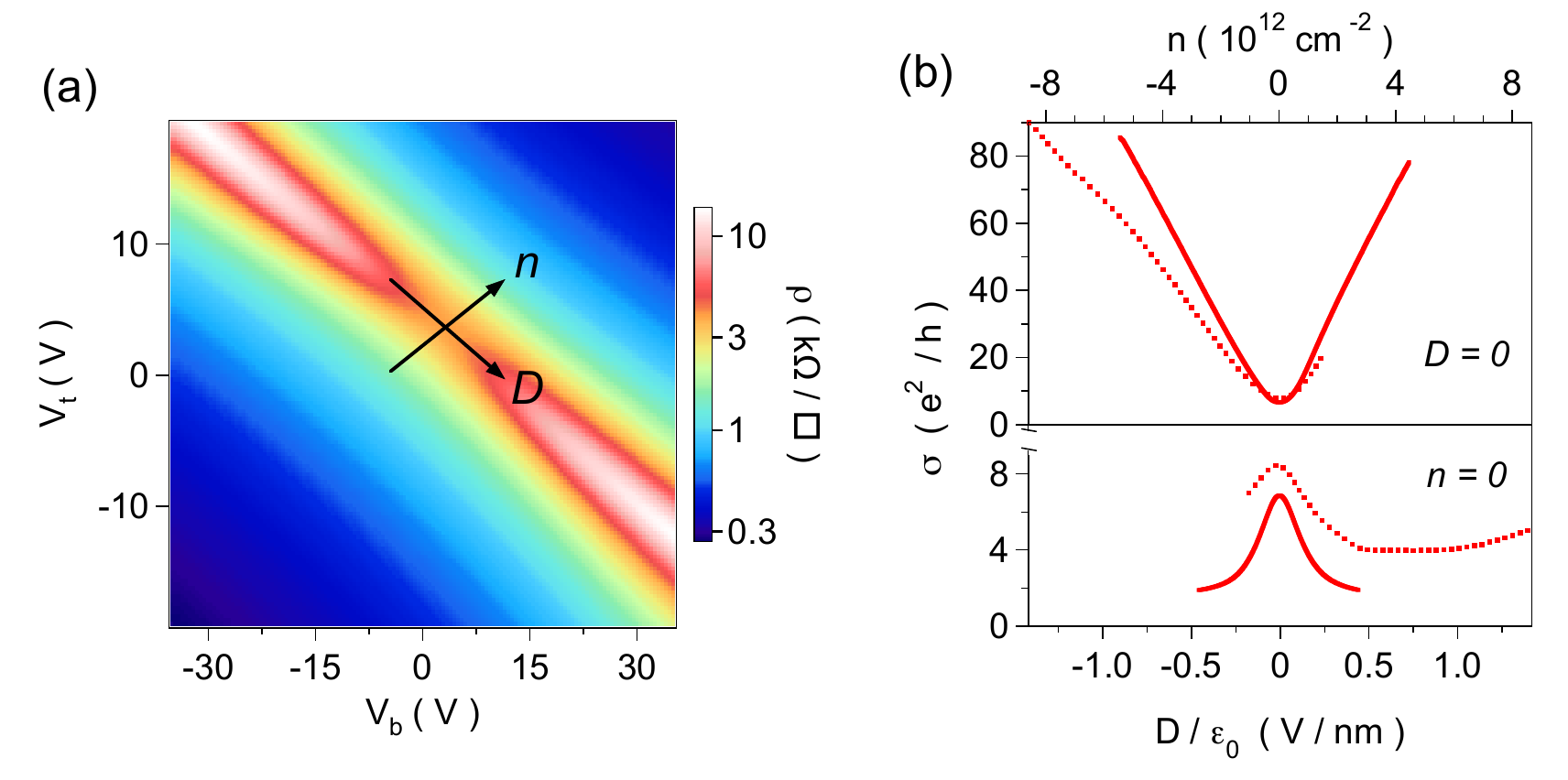}
\caption{(Color) (a) ABA trilayer sheet resistivity, $\rho$, in units of k$\Omega$ per square, vs. the top and back gate voltages, $V_\text{t}$ and $V_\text{b}$. The arrows at the saddle point indicate the axes of increasing carrier density, $n$, and electric displacement field, $D$.  (b) Profile cuts along the $n$ and $D$ axes are plotted as the conductivity, $\sigma$, vs. $n$ or $D$. The solid lines are from Fig. 2 (a), while the dashed lines are from a second sample in which the saddle point is offset from $V_\text{b}=V_\text{t}=0$ due to extrinsic doping, so that a larger range of $D$ is accessible.}
\end{figure*}

In Fig. 2 we show the sheet resistivity, $\rho$, at zero magnetic field, $B$, as a function of the gate voltages $V_\text{t}$ and $V_\text{b}$. Two arrows superimposed on the data define the axes of increasing $n$ and $D$ induced by the two gates; they cross at a saddle point in the resistivity which is independently identified as where $n=D=0$ via measurements at high magnetic field. This saddle point bears a superficial resemblance to the gate voltage dependence of the resistivity of bilayer graphene \cite{taychatanapat:166601,zou:081407}. Overall, the transport here is outlined via profile cuts along the $D$ and $n$ axes, plotted in Fig. 2 (b) as the conductivity, $\sigma$, against $n$ (top axis) or $D$ (bottom axis) respectively. Data from two samples are shown, with the solid lines taken from Fig. 2 (a), and the dotted lines from a second sample in which the saddle point was further displaced from $V_\text{b} = V_\text{t} = 0$ due to extrinsic doping, thereby allowing a greater range of $D$ to be accessed. As for monolayer and bilayer graphenes on SiO$_2$, $\sigma \propto n$ away from charge neutrality. At the saddle point, $\sigma_{\text{min}}= 7-8 ~e^2/h$ and is 2 to 3 times greater than for typical bilayer graphene on SiO$_2$, perhaps because the higher density of states in trilayer graphene leads to more effective screening of scattering sources \cite{xiao:041406,yuan:235409,min:195117}. The change in $\sigma$ with increasing $D$, however, is the most unique feature. Initially, $\sigma$ decreases by a factor of 2 or 3 until $D/\epsilon_0\approx 0.4$ V/nm, beyond which it is flat until, for $D/\epsilon_0>1.0$ V/nm, it begins to slowly rise again. This new and unusual behavior-- observed in multiple samples-- strongly suggests that, unlike bilayers, no gap opens in ABA trilayers at high $D$. This behavior is similar to predictions for the conductivity of ABA trilayers subject to an electric potential imbalance between the top and bottom graphene sheets, in which $\sigma(n=0)$ is generically expected to decrease sharply and then rise slowly as $D$ increases from zero \cite{koshino:125443}. 

We note that earlier work on dual-gated trilayer devices found entirely different behavior for the resistivity, showing an apparent global {\it maximum} at the point identified as $n=D=0$ \cite{craciun:383,jhang:4995}. However, the location of the $D=0$ point was not independently determined in those studies.

At high magnetic fields, the QHE for ABA-stacked trilayer graphene exhibits a number of properties that distinguish it from other graphenes. In Fig. 3 (a), the inverse of the measured Hall resistance, $R_{xy}^{-1}$, is plotted for $B = 14$ T. Several QH plateaus are clearly visible as wide stripes running from the upper left to lower right. Contour lines drawn at half-integer values of the filling factor, $\nu=n h / e B = ...-1.5,-0.5,+0.5,+1.5...$, serve to emphasize the boundary of each plateau so that, for instance, the $-6$ plateau is bounded by contours at $\nu=-6.5$ and $-5.5$. The interesting features of this data fall into three groups: First, the most obvious plateaus, running diagonally in parallel to the $D$ axis, comprise the sequence of quantum numbers $-14, -10, -6, +6,$ and $+10$ that are labeled in the figure. Second and most unusual, several distinct plateaus with quantum numbers $\pm2$ and $\pm4$ develop near the upper left and lower right corners of Fig. 3 (a), at low $n$ and high values of $D$. Finally, a small region at the center of the plot is found to be well quantized with a conductance of $-2e^2/h$. For all of these, the precise quantization of the plateaus is highlighted by cuts made along the $n$ ($D=0$) axis, as well as at the boundaries of Fig. 3 (a) at constant $V_\text{t}$ for varying $V_\text{b}$, and vice versa, all shown in Figs. 3 (b)-(d), respectively. 

Several of the QH features can be explained with recourse to the lowest-order model of the ABA trilayer band structure. In a tight-binding calculation when only terms describing the in-plane nearest-neighbor hopping and the largest interlayer hopping-- $\gamma_0$ and $\gamma_1$, respectively, in Fig. 1 (a)-- are included, the resulting band structure arising at the $K$ and $K'$ points in momentum space is depicted in Fig. 4 (a). It resembles a ``1+2'' superposition of linear and hyperbolic bands similar to those of monolayer and bilayer graphene \cite{guinea:245426,koshino:115315}. On the right of Fig. 4 (a) the corresponding zero-energy Landau levels (LLs) are schematically plotted as a function of position, so the electron or hole levels diverge upward or downward toward the sample edge. In monolayer (bilayer) graphene, one (two) fourfold degenerate LL(s), reflecting the spin and valley degrees of freedom, are found at zero energy. Thus, in this 1+2 model, three such zero-energy levels appear and give rise to an overall 12-fold degeneracy, yielding quantum numbers $+6$ or $-6$ when the Fermi level lies where indicated by the dashed lines. Consequently, accounting for additional higher-energy LLs will, barring accidental degeneracies, lead to the sequence $...-14,-10,-6,+6,+10,+14...$ which is identical to that observed along the line where $D=0$ {\it except} for the presence of the $-2$ plateau.

\begin{figure*}[t]
\includegraphics[width=0.9\textwidth]{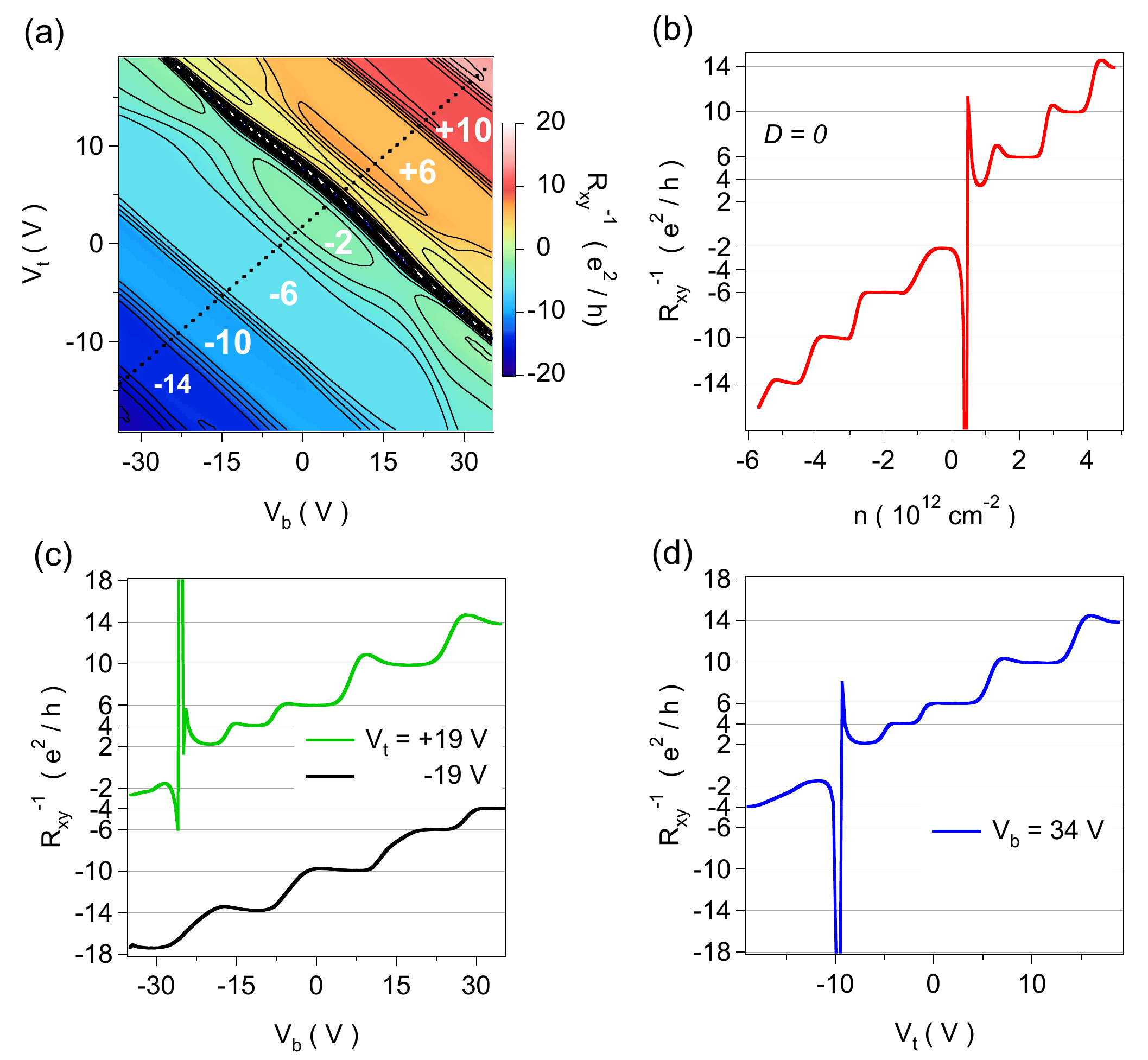}
\caption{(Color) (a) The ABA trilayer quantum Hall effect, plotted as the inverse Hall resistance, $R_{xy}^{-1}$, vs. $V_\text{b}$ and $V_\text{t}$ at $B=14$ T. Contour lines are drawn at filling factors $\nu = ...-1.5,-0.5,+0.5,+1.5...$. The quantum numbers of several plateaus are indicated in white. Several profile cuts through the data to highlight the various quantized plateaus are shown in (b), (c), and (d). The cut in (b) is taken along $D=0$ following the dashed line in (a). The cuts for (c), for constant top gate voltages, are taken along the top and bottom edges of (a). The cut in (d) is taken along the right edge of (a).}
\end{figure*}

The appearance of the $\pm2$ and $\pm4$ quantized plateaus in the upper left and lower right corners of Fig. 3 (a), highlighted in Fig. 3 (c) and (d), can also be understood in a lowest order model of ABA trilayers. In the theoretical picture of Ref.~\cite{koshino:115315}, the ABA trilayer Hamiltonian may be decomposed into one monolayerlike and one bilayerlike system, and since each of these alone has inversion symmetry, an effective inversion symmetry is recovered for the ABA trilayer. This guarantees the degeneracy of states at the $K$ and $K'$ valleys, despite the lack of a true lattice-inversion symmetry. However, this model does not hold in the presence of a potential drop between the graphene layers due, e.g., to the external $D$ field, and the valley degeneracy is thus lifted. In particular, when $D \neq 0$, the three fourfold degenerate LLs at zero energy split, so that one twofold degenerate level (for electron spin) from each valley remains at zero energy, while the remaining levels disperse with a dependence on $D$ that differs in each valley. Therefore, as $D$ increases from zero, these diverging LLs will initially give rise to new quantized plateaus in a 4,2,-2,-4 sequence \cite{koshino:115315,yuan:125455} that is strikingly similar to the pattern of the QH plateaus observed in Fig. 3. With further increase of $D$, additional plateaus are expected at even values of the filling factor, and indeed in a second sample the $\nu=+8$ and $+12$ plateaus have also been seen. Since only the strength, and not the sign, of the applied field is important, the 4,2,-2,-4 structure appears symmetrically for $\pm D$. Thus, this feature can be used as an independent method to locate the line where $D=0$, which is found to pass through the saddle point of Fig. 2 (a), justifying the previous identification of the saddle point as where $n=D=0$. Finally, we note that the QHE in ABC trilayers for devices of similar quality is expected to be distinctly different, with a filling factor sequence $\nu=...-6,0,+6...$ for any value of $D$ \cite{koshino:165409}.

While the lowest-order model of ABA trilayers successfully explains several of the QH plateaus in Fig. 3 (a), it offers no insight into the presence of the robust $-2$ plateau near the center of the plot, nor to the corresponding asymmetry reflected in the lack of a plateau at $\nu=+2$. Clearly, the $\nu=-2$ plateau must arise from a splitting of the three zero-energy LLs. While in principle many-body interactions can be responsible, in these low-mobility samples ($\mu\approx4000$ cm$^2$/Vs) disorder is expected to overwhelm such effects. Therefore, we seek an explanation for the $-2$ plateau in terms of a more accurate model of the trilayer band structure \cite{koshino:165443}. 

\begin{figure}[t]
\includegraphics[width=\columnwidth]{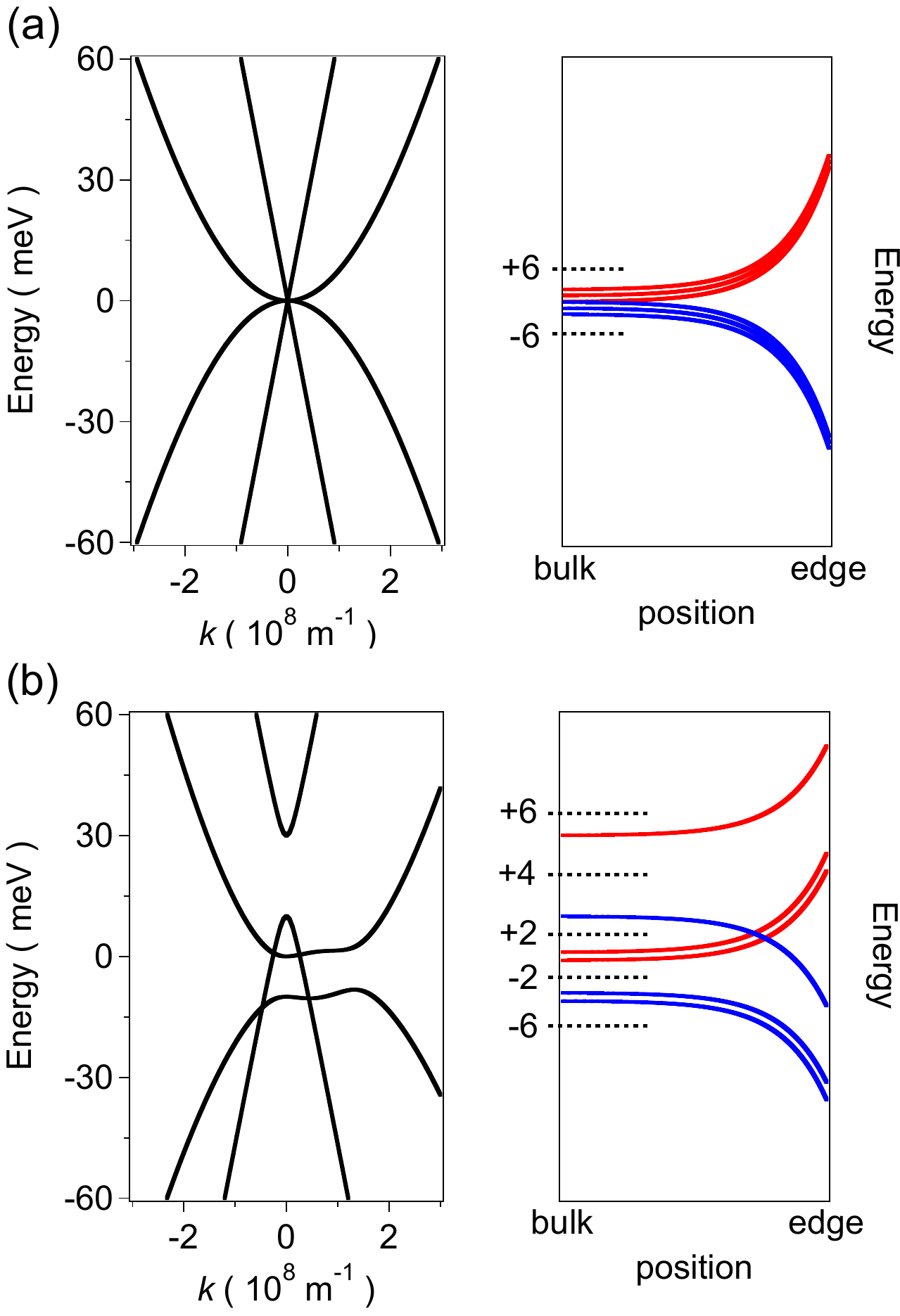} 
\caption{(Color) (a) The band structure of the ``1+2'' model of ABA trilayer graphene (left) and the associated lowest LLs (right). QH plateaus will occur when the Fermi level lies at the dashed lines, with quantum number $+6$ ($-6$) when all levels are full (empty). (b) Band structure calculated to higher order in the Slonczewski-Weiss-McClure model (see text). Two bands overlap at zero energy, creating a semimetal. To the right, the corresponding lowest LLs allow for additional QH plateaus at $-2, +2,$ and $+4$. Level energies are plotted against position, diverging upward (downward) for electron (hole) states near the sample boundary. All LLs are twofold degenerate for the electron spin. Closely spaced pairs only mark the number of spin-degenerate levels, not additional splittings.} 
\end{figure}

In Fig. 4 (b), the ABA band structure is shown, calculated to higher order by utilizing the full set of tight-binding parameters in the Slonczewski-Weiss-McClure model for graphite \cite{castroneto:109} with values identical to those used in Ref.~\cite{koshino:125443}: $v_0(\propto \gamma_0)=1.0\times10^6$ m/s, $\gamma_1=0.4$ eV, $\gamma_2=-0.05\gamma_1$, $v_3 (\propto \gamma_3) = 0.1v_0$, $v_4(\propto \gamma_4) = 0.014v_0$, $\gamma_5 = 0.1\gamma_1$, and $\delta=0.125\gamma_1$, as well as $D=0$. In this picture, the monolayerlike and bilayerlike bands become gapped as well as offset from zero energy, leading to the overlap of one approximately linear hole band with a nearly parabolic electron band. The region of overlap delimits the semimetal: At charge neutrality, {\it both} bands are partially occupied and contribute to transport. 

The lowest energy LLs for the higher order calculation are plotted on the right of Fig. 4 (b) as twofold degenerate levels (for electron spin), with one (two) arising at each extrema of the monolayerlike (bilayerlike) bands. Some surprising features appear: New gaps develop at $\nu=-2, +2,$ and $+4$, and a LL crossing naturally arises as the levels diverge at the sample edge. Thus, the higher order model can account for the observation of a plateau at $\nu=-2$. But why then are plateaus absent at $\nu=+2$ and $+4$? In fact, the relative sizes of the gaps shown at $\nu=-2, +2,$ and $+4$ in Fig. 4 (b) may not be correct: The gaps shown here depend sensitively on the values of the tight-binding parameters used in the higher-order calculation, and values of these are not well known for trilayer graphene. The data in Fig. 3 would suggest that the $\nu=-2$ gap is larger than those at $\nu=+2$ or $+4$, which may be washed out by disorder. However we note that a dip appears in $R_{xy}^{-1}$ near $n=1\times10^{12}$ cm$^{-2}$ in Fig. 3 (b), possibly signaling an incipient $+2$ or $+4$ plateau. Figure 4 (b) further illustrates an intriguing feature of the $\nu=+2$ state, due to the LL crossing of electron and hole bands which carry counter-propagating edge modes. While it is unclear whether this should affect the quantization, this feature may lead to new physics in future high mobility devices \cite{abanin:196806,fertig:116805}.

\begin{figure*}[t]
\includegraphics[width=0.9\textwidth]{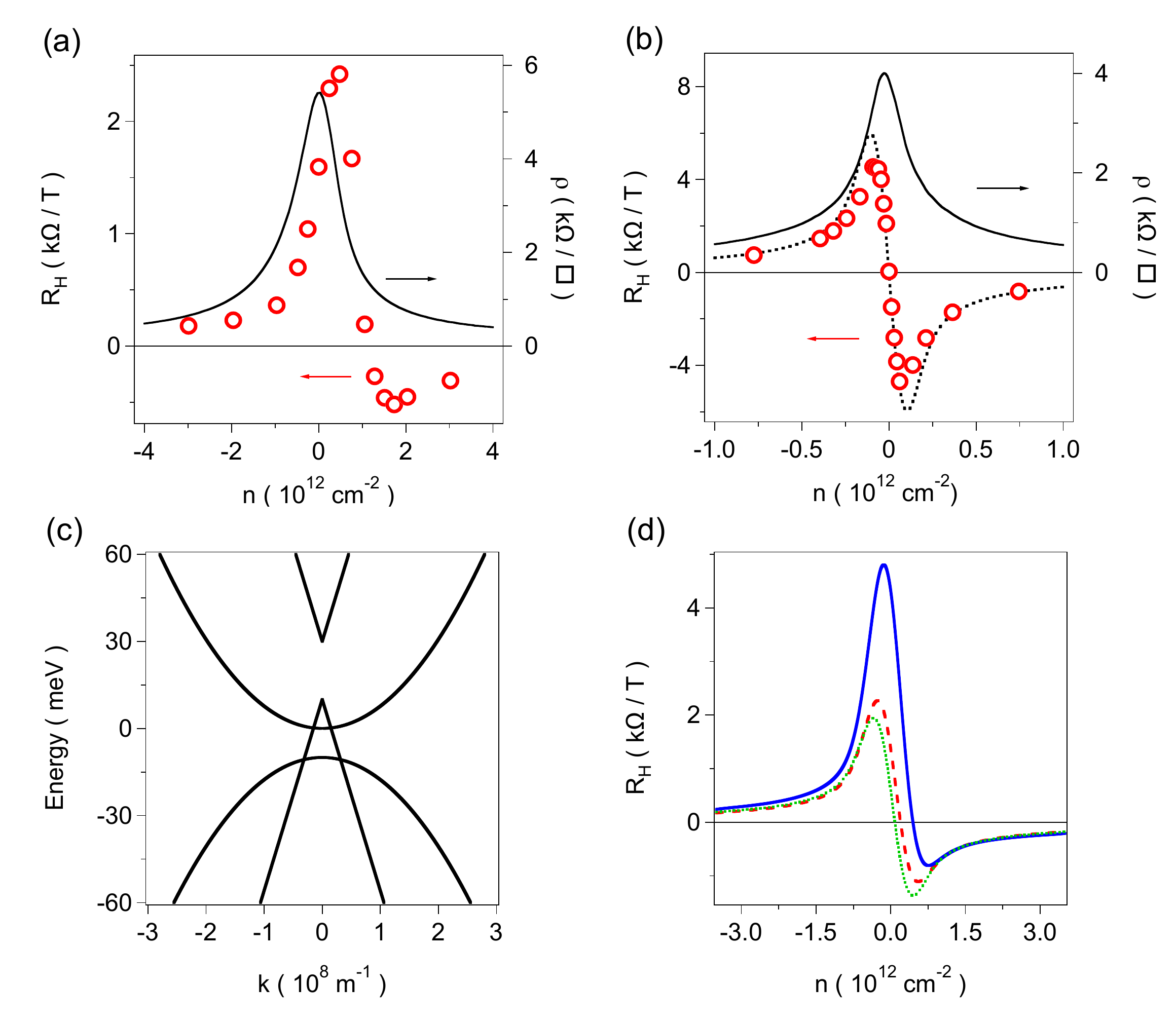} 
\caption{(Color) (a) The low field Hall coefficient, $R_\text{H}$, and sheet resistivity, $\rho$, in k$\Omega$ per square, measured in ABA trilayer graphene as a function of $n$, for $D=0$. Uncertainty in the measurements is smaller than the symbol size. (b) $R_\text{H}$ and $\rho$ measured in monolayer graphene. The dotted line is $R_\text{H}=-1/ne$, averaged over disorder (see text). (c) Simplified ABA band structure. (d) $R_\text{H}$ vs. $n$ calculated for the band structure in (c), for mobility ratios $\mu_\text{l}/\mu_\text{p} = 3.3$ (solid blue), $1$ (dashed red) and $0.3$ (dotted green) (see text).} 
\end{figure*}

As the LL sequence and electron-hole asymmetry in the center of Fig. 3 (a) are both consistent with a semimetallic band overlap in ABA trilayers, it is natural to assume that low-field measurements will also show semimetallic transport. To address this assumption, the low-field Hall coefficient, $R_\text{H} = dR_{xy}/dB$, and sheet resistivity, $\rho$, are compared to a control sample of monolayer graphene. The results are shown in Fig. 5 (a) and (b) for trilayer and monolayer samples, respectively, where the measured $R_\text{H}$ and $\rho$ are plotted against carrier density. The trilayer data are acquired along $D=0$, which is possible only in dual-gated samples. The evolution of $R_\text{H}$ is quite different for the two systems. The monolayer data are perfectly symmetric about $n=0$; due to disorder, they do not diverge but rather smoothly change sign over a narrow region at low densities. In contrast, the trilayer data are strongly asymmetric, having a large positive peak at $n=0$ and a zero crossing at slightly more than $n=1\times10^{12}$ cm$^{-2}$, followed by a weak minimum and slow decrease back toward zero. Meanwhile, in ABA trilayers, a $\rho_{xx}\sim B^2$ behavior is observed, characteristic of two-band conduction. This observation, along with analysis of the clear two-band signatures in the Shubnikov-de Haas oscillations, will be discussed in future work.

The different behavior of $R_\text{H}$ can be qualitatively understood in terms of semimetallic transport. As expected for conduction arising from a single band, the monolayer data are well described by $R_\text{H} = -1/ne$ except at the lowest densities, which are modeled by convolving $R_\text{H}$ with a Gaussian whose variance, $\delta n$, represents a spread of densities due, e.g., to electron-hole puddles \cite{martin:144,henriksen:041412}. While this model certainly oversimplifies the role of disorder, a reasonable fit to the data can be found for $\delta n=0.8 \times 10^{11}$ cm$^{-2}$ (dashed line in Fig. 5 (b) ). In contrast, the transport properties are qualitatively different when two bands take part in the conduction, as occurs in ABA trilayers. In particular, when electron and hole bands overlap, the low-field Hall coefficient becomes \cite{beer}
\begin{equation}
R_\text{H} ~=~ \frac{1}{e}~\frac{p - n\left(\mu_\text{e}/\mu_\text{h}\right)^2}{\left(p+n \mu_\text{e}/\mu_\text{h}\right)^2},
\end{equation}
where $p$ and $n$ are the hole and electron densities and $\mu_\text{e}/\mu_\text{h}$ is the ratio of their mobilities. Sharply differing from the behavior for a single band, this two-band form of $R_\text{H}$ will in general cross zero away from charge neutrality ($p=n$); and at the zero-crossing, the ratio of the densities is given by $p/n=\left(\mu_\text{e}/\mu_\text{h}\right)^2$.

In Fig. 4 (b), the number of conducting bands-- and associated mobilities-- changes several times as the Fermi level is varied. To model $R_\text{H}^\text{tri}$, the simplified ABA band structure shown in Fig. 5 (c) is adopted. Depending on the Fermi level position, the relevant one- or two-band expression is employed, and the result is smoothed with a Gaussian using $\delta n = 3\times 10^{11}$ cm$^{-2}$. Obviously, these results will vary with the choice of mobilities and band offsets, but overall the behavior of this model can be summarized by the curves in Fig. 5 (d). There, $R_\text{H}$ vs. $n$ is calculated for three ratios of the band mobilities, $\mu_\text{l}/\mu_\text{p} = 3.3, 1$, and $0.3$, where $\mu_\text{l}$ and $\mu_\text{p}$ are mobilities of the linear and parabolic bands, irrespective of the sign of the carriers. Interestingly, the data are best approximated when $\mu_\text{l} > \mu_\text{p}$, which gives the only curve (blue line in Fig. 5 (d) ) that captures the pronounced asymmetry of the data and crosses zero at a positive carrier density. Of course, the many choices of band-edge offsets and curvatures, combined with the likely density-dependent mobilities, all prohibit using this model to make accurate fits to the data. In particular, substrate-induced potential fluctuations are ubiquitous in SiO$_2$-based devices and likely to contribute to a smearing of transport effects over a range of densities \cite{hwang:081409,dassarma:407}. Nonetheless, the simple semimetallic model of Fig. 5 (c) does capture the essence of the $R_\text{H}^\text{tri}$ data. These results, in concert with the clear asymmetry of the quantum Hall effect, strongly suggest that ABA trilayer is indeed a semimetal with an unusual LL structure at low energies.

\bigskip

\noindent{\bf Acknowledgments:} We gratefully acknowledge conversations with S. Adam, J. Alicea, Y. Barlas, D. Bergman, B. Chickering, K. C. Fong, M. Koshino, C. H. Lui, A. MacDonald, and E. McCann. Special thanks are due G. Rossman for the use of his spectroscopy lab. This work is supported by the DOE under grant No. DE-FG03-99ER45766 and the Gordon and Betty Moore Foundation.

\bibliographystyle{Science}

\end{document}